\documentclass[journal=jpcafh,manuscript=article]{achemso}
\setkeys{acs}{usetitle=true}
\usepackage[version=3]{mhchem} 
\usepackage{multirow}
\usepackage{amsmath, amsthm, amssymb}
\usepackage{natmove}
\usepackage{achemso}
\usepackage{graphicx}
\usepackage{bm}				
\author{Caterina Cocchi}
\email{caterina.cocchi@physik.hu-berlin.de}
\affiliation{Centro S3, CNR-Istituto Nanoscienze, I-41125 Modena, Italy}
\alsoaffiliation{Dipartimento di Scienze Fisiche, Informatiche e Matematiche, Universit\`a di Modena e Reggio Emilia, I-41125 Modena, Italy}
\altaffiliation{Present address: Humboldt-Universit\"at zu Berlin, Institut f\"ur Physik und IRIS Adlershof, Zum Grossen Windkanal 6, 12489 Berlin, Germany; e-mail: caterina.cocchi@physik.hu-berlin.de}
\author{Deborah Prezzi}
\affiliation{Centro S3, CNR-Istituto Nanoscienze, I-41125 Modena, Italy}
\author{Alice Ruini}
\affiliation{Centro S3, CNR-Istituto Nanoscienze, I-41125 Modena, Italy}
\alsoaffiliation{Dipartimento di Scienze Fisiche, Informatiche e Matematiche, Universit\`a di Modena e Reggio Emilia, I-41125 Modena, Italy}
\author{Marilia J. Caldas}
\affiliation{Instituto de F{\'\i}sica, Universidade de S\~ao Paulo, 05508-900 S\~ao Paulo, SP, Brazil}
\author{Elisa Molinari}
\affiliation{Centro S3, CNR-Istituto Nanoscienze, I-41125 Modena, Italy}
\alsoaffiliation{Dipartimento di Scienze Fisiche, Informatiche e Matematiche, Universit\`a di Modena e Reggio Emilia, I-41125 Modena, Italy}
\title{ Anisotropy and Size Effects on the Optical Spectra of Polycyclic Aromatic Hydrocarbons}
\begin{document}
\newpage
\begin{abstract}
The electronic and optical properties of polycyclic aromatic hydrocarbons (PAHs) present a strong 
dependence on their size and geometry.
We tackle this issue by analyzing the spectral features of two prototypical classes of PAHs, 
belonging to $D_{6h}$ and $D_{2h}$ symmetry point groups and related to coronene as multifunctional seed.
While the size variation induces an overall red shift of the spectra and a redistribution 
of the oscillator strength between the main peaks, a lower molecular symmetry is responsible 
for the appearance of new optical features. 
Along with broken molecular orbital degeneracies, optical peaks split and dark states are activated in the low-energy part of the spectrum.
Supported by a systematic analysis of the composition and the character of the optical transitions, 
our results contribute in shedding light to the mechanisms responsible for spectral modifications 
in the visible and near UV absorption bands of medium-size PAHs.
\end{abstract}
\newpage
Polycyclic aromatic hydrocarbons (PAHs) are the focus of research in significantly diverse fields, 
ranging from biochemistry \cite{arfs+96ees} to theoretical chemistry \cite{clar-scho64book}, from molecular electronics \cite{wu+07cr} to astrochemistry \cite{tiel08araa}, where PAHs are assigned an important role \cite{lege-puge84aa,alla+85apj} due to their chemical composition and spectral features \cite{lege-dhen85aa,vand-alla85aa,sala+99apj,alla+89apjs,alla+99apjl}.
In this framework, a theoretical identification of the absorption and emission features of these molecules assumes a significant relevance \cite{mall+07cp,baus+10apjs}, since they might drastically vary both upon small structural modifications \cite{mall+04aa,rieg-muel10jpoc} and as a consequence of ionization \cite{craw+85apj,canu+91apj,rast+13aipcp,care+13aa} or dehydrogenation \cite{mall+08aa,care+10apj,care+11mnras,baus-ricc13apj}.
In the field of molecular electronics, the interest in medium- and large-sized PAHs is intrinsically connected to graphene, as they are viewed as highly stable graphene molecules \cite{wats+01cr}.
Their intriguing and versatile optical properties \cite{doet+00jacs,liu+11jacs}, combined with their facile bottom-up synthesis \cite{zhi-muel08jma} and their capability to assemble into $\pi$-$\pi$ stacked superstructures \cite{feng+09pac,pisu+10cm}, make them excellent candidates for opto-electronic applications, \cite{grim-muel05ac,wu+07cr} where slight structural variations can trigger different properties and functionalities \cite{pisu+09mrc}.

In this paper we investigate the effects of size and anisotropy  on the optical spectra of PAHs, with the aim of providing an useful reference for understanding and predicting their optical response.
To tackle this issue, we focus on two prototypical classes of molecules with $D_{6h}$ and $D_{2h}$ point group symmetry, which are related to coronene, viewed as a multifunctional seed.
Indeed, with its high symmetry and specific electronic structure, coronene is one of the most representative medium-size PAHs \cite{wats+01cr}.
By means of a well-established quantum chemistry approach, we perform a theoretical  analysis to identify the size- and anisotropy-induced modifications to the spectral features in the visible and near-UV bands of PAHs.
The present study aims at showing that the main effect of size increase, moving from coronene to larger hexagonal PAHs of fixed symmetry, is an oscillator strength variation of the main excitations and a red shift of the whole spectrum, which brings the strong absorption peaks from the UV to the visible energy region, maintaining overall the character of the excitons.
Moreover, the additional features produced in the spectra by the anisotropical elongation of coronene in one direction will be investigated; specific focus will be devoted to the effects of symmetry breaking.
Finally, it will be clarified that bright excitations in the visible region appear already for the smaller anisotropic molecules, with low mass compared to highly symmetric PAHs.

%
\section{Computational Methods}
%
The results presented in this paper are obtained within the framework of Hartree-Fock based semi-empirical methods \bibnote{AM1 and ZINDO/S calculations were performed using VAMP package included in Accelrys Materials Studio software, version 5.0  (\url{http://accelrys.com/products/materials-studio}).}.
This methodology is well tested and reliable to compute the electronic and optical properties of C-conjugated low-dimensional systems 
\cite{wetm+00cpl,cald+01apl, davi-cald02jcc,kuba+02nat}, including PAHs themselves \cite{canu+91apj,steg+10apjl}.
The AM1 model \cite{dewa+85jacs} is adopted for structural optimization 
[0.4 $\text{kcal}$/(mol $\cdot$ \AA{})
threshold for the forces], whereas the ZINDO/S model \cite{ridl-zern73tca}, based on Configuration Interaction (CI) including single excitations only, is employed for the optical spectra.

To better characterize the optical excitations, we introduce the \textit{transition density} 
\cite{fisc71jmsp,dreu-head05cr}, which is defined as:
\begin{equation}
\rho^I (\mathbf{r}) = \sum_{\alpha,\beta} c_{\alpha \beta}^I \phi_{\beta}^*(\mathbf{r}) \phi_{\alpha}(\mathbf{r}) ,
\label{eq:td}
\end{equation}
where $\phi_{\alpha}$ ($\phi_{\beta}$) indicates the occupied (unoccupied) 
molecular orbitals (MOs) and $c^I_{\alpha \beta}$ are the CI coefficients of the $I^{th}$ excited-state configuration \cite{jens07book}.
The transition density is directly related to the transition dipole moment of the $I^{th}$ excitation \cite{been-pull04jpcb}:
\begin{equation}
\label{eq:trans-dip}
\mathbf{\mu}^I_{0p} =  \langle \Phi_0|\widehat{\mathbf{p}}|\Phi_p \rangle \: = -e \int d^3r \: \widehat{\mathbf{r}} \: \rho^I(\mathbf{r}) ,
\end{equation}
where $\widehat{\mathbf{p}} = -e \widehat{\mathbf{r}}$ is the momentum operator, $\Phi_0$ is the ground state Hartree-Fock wave function and $\Phi_p$ the singly excited configuration, including all the possible Slater determinants obtained by coupling $\phi_{\alpha} \rightarrow \phi_{\beta}$ transitions in the given energy window.
From \ref{eq:trans-dip} it is thus clear that the transition density provides a representation of the polarization of the excitation, which can be visualized through a three dimensional isosurface representation \cite{krue+98jpcb,sun+05cpl}.
Finally, as the oscillator strength is proportional to the square modulus of the transition dipole moment, the transition density also carries information about the excitation intensity.

%
\begin{figure}
\centering
\includegraphics[width=.9\textwidth]{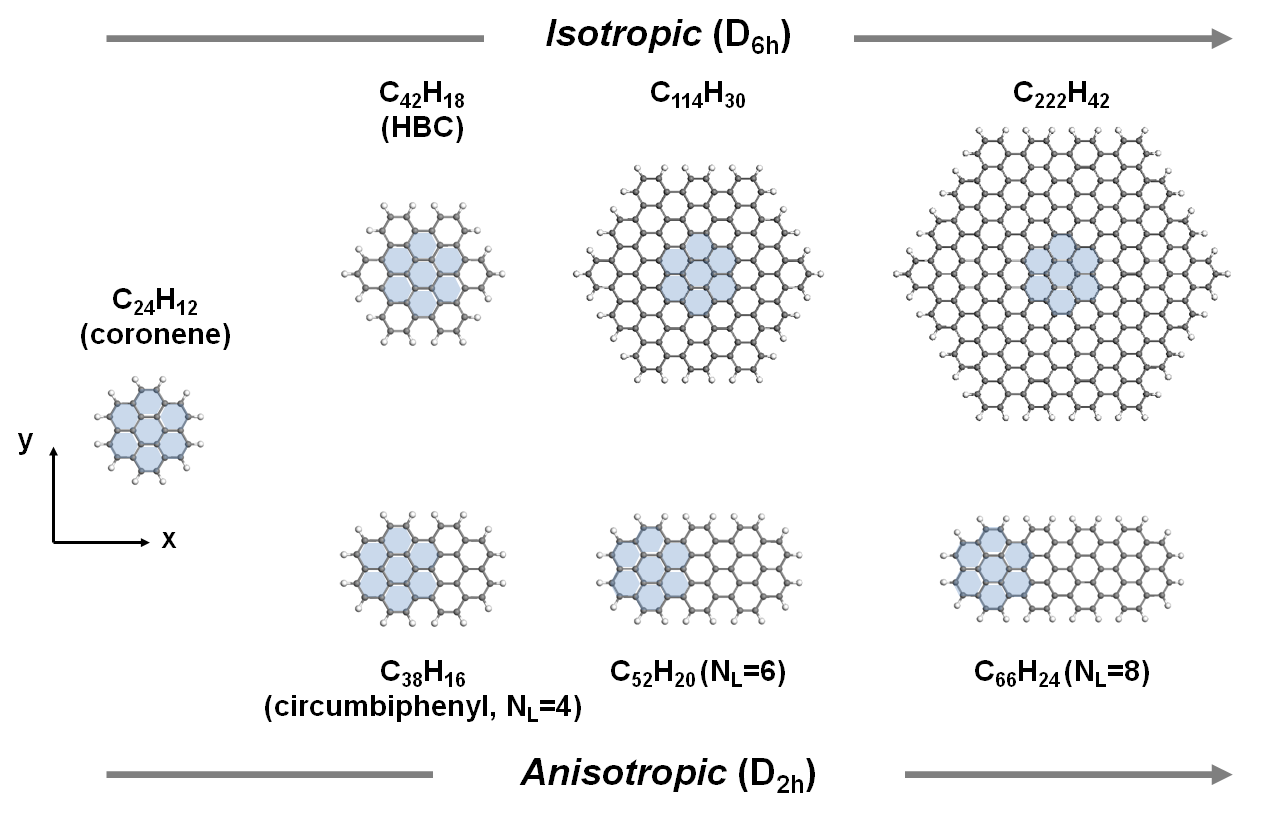}%
\caption{Polycyclic aromatic hydrocarbons (PAHs) investigated in this work. 
Starting from a coronene molecule as multifunctional seed, we consider both 
its isotropic and anisotropic size increase. In the first case (above) we obtain hexagonal 
PAHs with $D_{6h}$ point group symmetry, while in the latter case (below) the hexagonal 
symmetry is broken and elongated PAHs, with $D_{2h}$ point group symmetry 
are formed. 
}
\label{fig1}
\end{figure}
Starting from coronene (\ce{C24H12}) as multifunctional seed, 
the effects of size and symmetry modifications on the optical properties are investigated by comparing two families of PAHs, characterized by $D_{6h}$ and $D_{2h}$ point group symmetry (see \ref{fig1}).
The first class of hexagonal molecules is a Kekul\'e series of fully-benzenoid PAHs with purely armchair edges \cite{stei-brow87jacs}, obtained by progressively incrementing the number of aromatic rings at the periphery.
As a result, the molecules belonging to this group are characterized by an increasing number of armchair-shaped rings per edge, ranging from 2 in hexa-benzo-coronene (HBC, \ce{C42H18}) up to  4 in \ce{C222H42} (see \ref{fig1}).
The second class of molecules is built by anisotropically elongating coronene 
in one direction only ($x$ axis), as indicated in \ref{fig1}. 
This gives rise to ribbon-like graphene nanoflakes (GNFs) with armchair-shaped edges.
These structures belong to  the $D_{2h}$ point group and can be identified through their length ($N_L$) and width parameters ($N_W$) \cite{cocc+12jpcl}.
The latter is defined as for periodic armchair graphene nanoribbons as the number of C-dimers along the $y$ axis of the molecule (see \ref{fig1}) \cite{naka+96prb}, while $N_L$ is identified as the number of zigzag chains in the $x$ direction, excluding the ends.
The coronene-derived GNFs considered here have fixed width of about 7.4 \AA{} ($N_W$=7) and increasing length, ranging from about 11 \AA{} ($N_L$=4, circumbiphenyl) to almost 20 \AA{} ($N_L$=8).
Note that in both cases we focus on fully armchair-edged molecules to avoid any spurious open-shell effects, typically associated with the presence of zigzag edges \cite{sun-wu12jmc}.

\section{Results and Discussion}
\begin{figure}
\centering
\includegraphics[width=.45\textwidth]{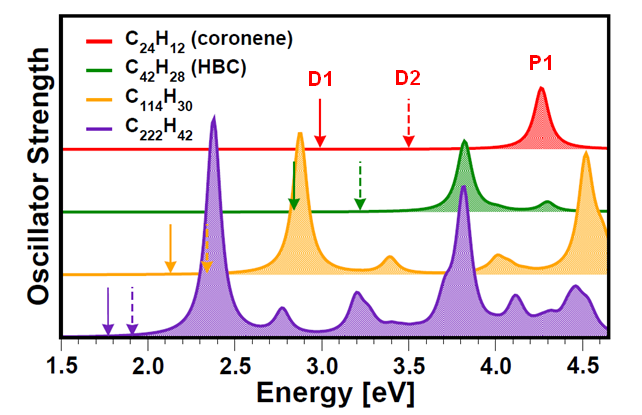}%
\caption{Calculated optical absorption spectra of hexagonal PAHs ($D_{6h}$ point group symmetry) of different sizes. The first intense 
peak P1 (Clar's $\beta$ band) is given by doubly degenerate excitations. 
The lowest energy dipole-forbidden excitations (D1 and D2, $\alpha$ and $p$ bands, according to Clar's notation, see also \ref{table1}) are indicated in the spectra by solid and dashed arrows, respectively. The spectra are obtained by using a Lorentzian broadening of 100 meV.
}
\label{fig2}
\end{figure}
%

The optical absorption spectrum of coronene (\ref{fig2}, top panel) is characterized by an intense peak in the near-UV region (P1, at 4.26 eV, see \ref{table1}) and by two low energy dark excitations in the visible band (D1 and D2 at about 3 and 3.5 eV, respectively).
The lowest energy excited states of hexagonal PAHs, corresponding to transitions between the doubly degenerate frontier orbitals (see \ref{fig4}a) and here labeled as D1, D2 and P1, are commonly identified as $\alpha$, $para$ ($p$) and $\beta$ bands, respectively, according to Clar's notation (see e.g. Ref. \cite{rieg-muel10jpoc} and references therein). 
These excitations correspond to symmetry forbidden $B_{1u}$ and $B_{2u}$ states (D1 and D2 in \ref{fig2}), and to a higher energy $E_{1u}$ state, characterized by a twofold degeneracy and by a large oscillator strength (P1 peak).

Within the CI approach, we find that P1 stems from the combined transitions of doubly degenerate frontier orbitals (highest occupied MO, HOMO, and lowest unoccupied MO, LUMO, see \ref{table1}).
Also D1 and D2  arise from frontier orbital transitions that combine here with opposite sign, hence giving rise to dipole-forbidden transitions. 
Two additional excitations -- one of them doubly degenerate -- appear in the spectrum between D2 and P1: they are symmetry-forbidden and come from combinations of transitions between the four occupied and unoccupied MOs closest to the gap.

\begin{figure}
\centering
\includegraphics[width=.95\textwidth]{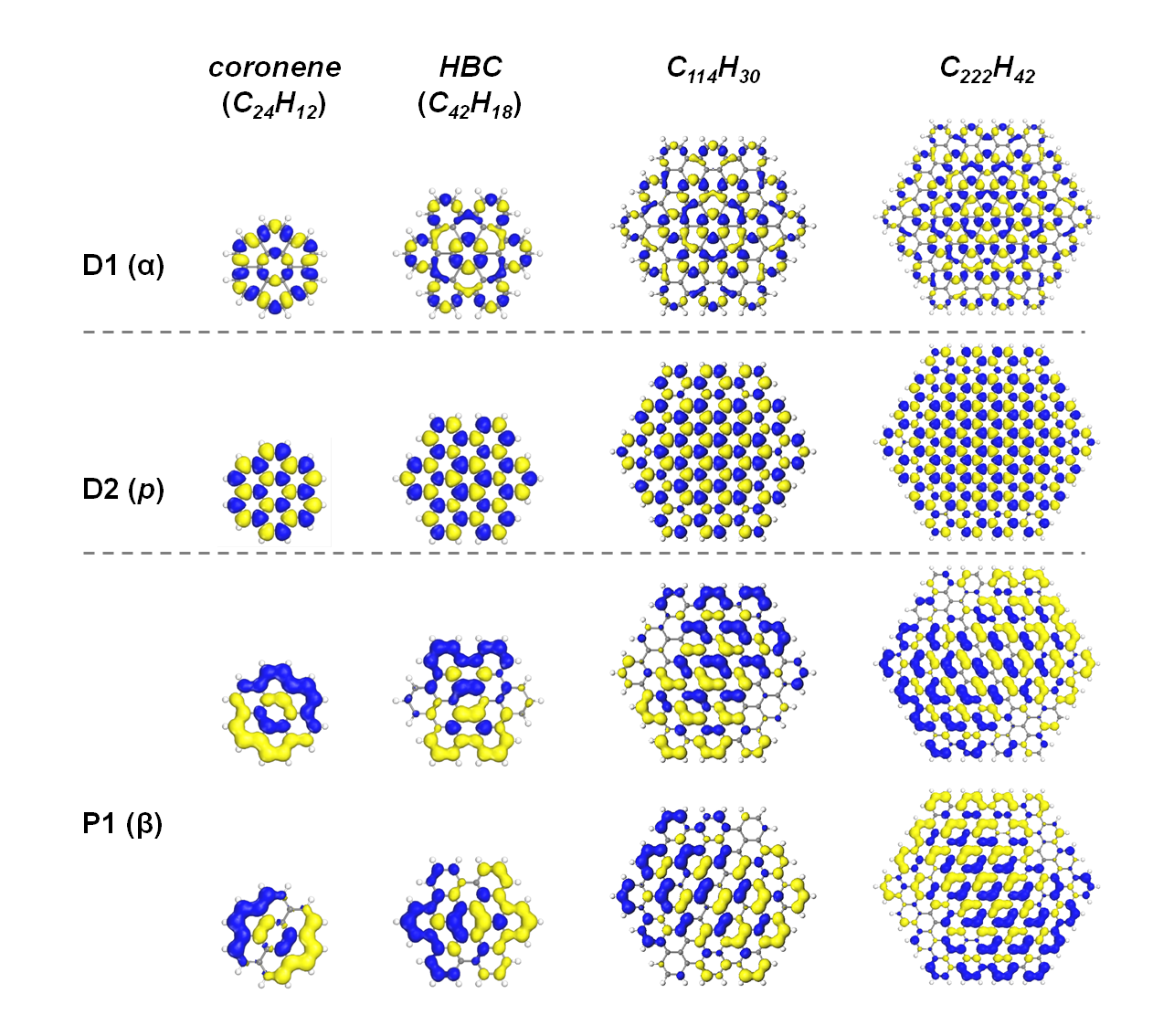}%
\caption{Transition density isosurfaces, computed as in \ref{eq:td}, of the examined optical excitations of hexagonal PAHs ($D_{6h}$ point group symmetry).}
\label{fig3}
\end{figure}

\begin{figure}
\centering
\includegraphics[width=.95\textwidth]{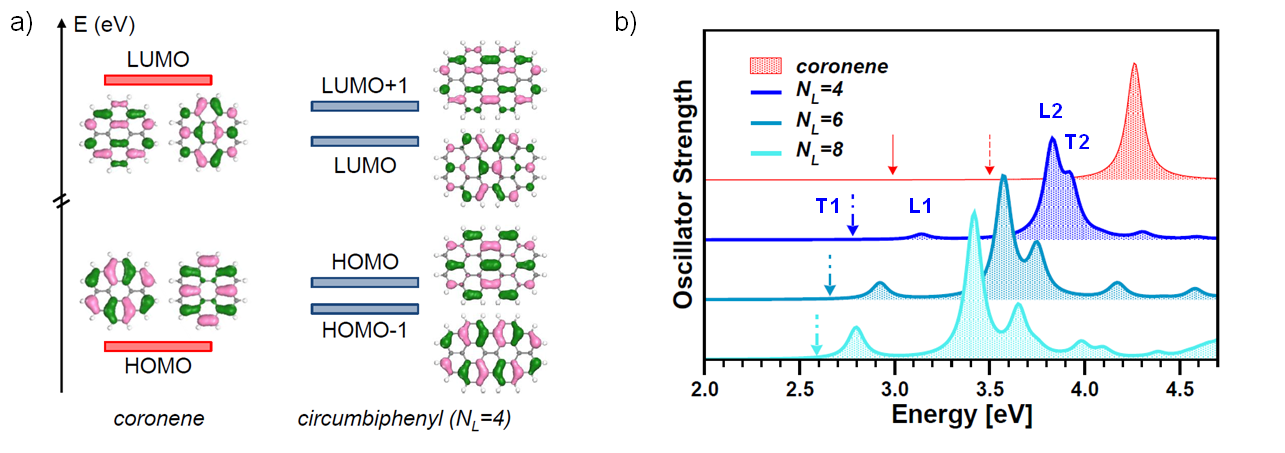}%
\caption{a) Isosurfaces of the frontier molecular orbitals of coronene and circumbiphenyl (ribbon-like PAH with $N_L$=4). The degeneracy of HOMO and LUMO of coronene is broken in the $D_{2h}$ PAHs. b) Calculated optical absorption spectra of ribbon-like PAHs, with fixed width $N_W$=7 (about 7.4 \AA{}) and variable length parameter ranging form $N_L$=4 to $N_L$=8. The main optical excitations, polarized either in the longitudinal (L, $x$ axis) or transverse (T, $y$ axis) direction are indicated in the spectrum of circumbiphenyl ($N_L$=4). The dashed-dotted arrows indicate the position of the lowest energy, weak excitation T1. For comparison, the spectrum of coronene is also shown (shaded red curve, top, see also \ref{fig2}). 
The spectra are obtained through a Lorentzian broadening of 100 meV.}
\label{fig4}
\end{figure}

The features of D1, D2 and P1 can be clearly captured from the transition density plots in \ref{fig3}.
The modulation of the positive and negative domains in D1 and D2 provides evidence of the zero dipole moment associated to them, while P1 is characterized by a remarkable polarity. 
These results are in agreement with the available experimental data \cite{katr+79cp,rieg-muel10jpoc} within about 150 meV, as well as with other predictions obtained within different theoretical approaches
 \cite{canu+91apj,weis+03apj,mall+04aa}.
The lowest energy peak is typically identified in experiments as the strongest absorption peak in the UV/vis region.
Moreover, as a matter of fact, the lowest dipole-forbidden excitations are also detected in experiments \cite{maru-iwas74cpl} as associated to vibrational modes \cite{orla-zerb88cp,canu+91apj} or, in general, with structural deviations from the ideal symmetry, generally adopted in theoretical simulations.
\bibnote{We recently addressed this issue in the analysis of the optical 
spectrum of icosahedral \ce{C60}. A detailed discussion can be found in Ref. \cite{cocc+13jpcc}.}.

\begin{table}
\begin{tabular}{c|c|c|c|c|c}
\hline  \hline
Molecule & Excitation & MO Transitions (weight) & $c_{\alpha \beta}^I$ & Energy [eV] & OS \\ \hline \hline
\multirow{8}{*}{\ce{C24H12}} & \multirow{2}{*}{D1 ($\alpha$)} & H $\rightarrow$ L+1 (0.47) & -0.683 & \multirow{2}{*}{2.99} & \multirow{2}{*}{-} \\ 
 & & H-1 $\rightarrow$ L (0.47) & 0.683 & & \\ \cline{2-6}
 & \multirow{2}{*}{D2 ($p$)} & H $\rightarrow$ L (0.49) & -0.700 & \multirow{2}{*}{3.50} & \multirow{2}{*}{-} \\
  & & H-1 $\rightarrow$ L+1 (0.49) &  -0.700 & & \\ \cline{2-6}
 & \multirow{4}{*}{P1 ($\beta$)} & H $\rightarrow$ L (0.48) & 0.690 & \multirow{4}{*}{4.26} & \multirow{4}{*}{1.96} \\
  & & H-1 $\rightarrow$ L+1 (0.48) & -0.690 & & \\ \cline{3-4}
 & & H $\rightarrow$ L+1 (0.48) &  0.690 & & \\
  & & H-1 $\rightarrow$ L (0.48) & 0.690 & & \\ \cline{3-4}
  \hline \hline
\multirow{12}{*}{\ce{C42H18}} & \multirow{2}{*}{D1 ($\alpha$)} & H $\rightarrow$ L+1 (0.44) & -0.665 & \multirow{2}{*}{2.84} & \multirow{2}{*}{-} \\ 
 & & H-1 $\rightarrow$ L (0.44) & 0.665 & & \\ \cline{2-6}
 & \multirow{2}{*}{D2 ($p$)} & H $\rightarrow$ L (0.47) & -0.684 & \multirow{2}{*}{3.22} & \multirow{2}{*}{-} \\
  & & H-1 $\rightarrow$ L+1 (0.47) & -0.684 & & \\ \cline{2-6}
 & \multirow{8}{*}{P1 ($\beta$)} & H $\rightarrow$ L (0.18) & 0.428 & \multirow{8}{*}{3.82} & \multirow{8}{*}{2.25} \\
  && H $\rightarrow$ L+1 (0.26) & -0.514 && \\
   & & H-1 $\rightarrow$ L (0.26) & -0.514 & & \\ 
  & & H-1 $\rightarrow$ L+1 (0.18) & -0.428 & & \\ \cline{3-4}
 & & H $\rightarrow$ L+1 (0.26) & -0.514 & & \\
   && H $\rightarrow$ L (0.18) & -0.428 && \\
   & & H-1 $\rightarrow$ L (0.18) & -0.428 & & \\ 
  & & H-1 $\rightarrow$ L+1 (0.26) & 0.514 & & \\ \cline{3-4}
 \hline \hline
\multirow{12}{*}{\ce{C114H30}} & \multirow{2}{*}{D1 ($\alpha$)} & H $\rightarrow$ L+1(0.42) & 0.649 & \multirow{2}{*}{2.13} & \multirow{2}{*}{-} \\  
 & & H-1$\rightarrow$ L (0.42) & -0.649 & & \\ \cline{2-6}
& \multirow{2}{*}{D2 ($p$)} & H $\rightarrow$ L (0.44) & 0.666 & \multirow{2}{*}{2.34} & \multirow{2}{*}{-} \\
  & & H-1 $\rightarrow$ L+1 (0.44) & 0.666 & & \\ \cline{2-6}
 & \multirow{8}{*}{P1 ($\beta$)} & H $\rightarrow$ L (0.41) & 0.638 & \multirow{8}{*}{2.88} & \multirow{8}{*}{4.53} \\
  && H $\rightarrow$ L+1 (0.05) & 0.228 && \\
   & & H-1 $\rightarrow$ L (0.05) & 0.228 & & \\ 
  & & H-1 $\rightarrow$ L+1 (0.41) & -0.638 & & \\ \cline{3-4}
 & & H $\rightarrow$ L+1 (0.41) & -0.638 & & \\
   && H $\rightarrow$ L (0.05) & 0.228 && \\
   & & H-1 $\rightarrow$ L (0.41) & -0.638 & & \\ 
  & & H-1 $\rightarrow$ L+1 (0.05) & -0.228 & & \\ \cline{3-4}
 \hline \hline
\multirow{8}{*}{\ce{C222H42}} & \multirow{2}{*}{D1 ($\alpha$)} & H $\rightarrow$ L+1 (0.40) & 0.629 & \multirow{2}{*}{1.77} & \multirow{2}{*}{-} \\ 
 & & H-1 $\rightarrow$ L (0.40) & -0.629 & & \\ \cline{2-6}
 & \multirow{2}{*}{D2 ($p$)} & H $\rightarrow$ L (0.42) & 0.646 & \multirow{2}{*}{1.91} & \multirow{2}{*}{-} \\
  & & H-1 $\rightarrow$ L+1 (0.42) & 0.646 & & \\ \cline{2-6}
 & \multirow{4}{*}{P1 ($\beta$)} & H $\rightarrow$ L (0.40) & 0.635 & \multirow{4}{*}{2.38} & \multirow{4}{*}{6.65} \\
  & & H-1 $\rightarrow$ L+1 (0.40) & -0.635 & & \\ \cline{3-4}
 & & H $\rightarrow$ L+1 (0.40) & 0.635 & & \\
  & & H-1 $\rightarrow$ L (0.40) & 0.635 & & \\ \cline{3-4}
  \hline \hline
\end{tabular} 
\caption{
Summary of the optical excitations of hexagonal PAHs ($D_{6h}$ point group symmetry) \textbf{indicated in the spectra of \ref{fig2}}, including energy, oscillator strength (OS) and composition in terms of molecular orbital transitions with the corresponding CI coefficients and weight. Only contributions larger than 5$\%$ have been included.
}
\label{table1}
\end{table}

For this series of hexagonal PAHs, \ref{fig2}, from the top to the bottom, indicates that the full optical spectrum, including dark excitations, is red-shifted with increasing size, as expected from the higher delocalization of the relevant molecular orbitals 
\bibnote{An exhaustive and systematic analysis of the size effects on the electronic and optical properties of a representative set of PAHs has been recently presented by Malloci and coworkers, see Ref. \cite{mall+11cp}.}, 
as recently reported for other PAH series \cite{rieg-muel10jpoc}.
Specifically, P1 is brought from the UV to the visible range (green), while the lowest-energy dark excitations move from the visible (violet) to the infrared band.
As such, P1 undergoes a larger red shift than D1 and D2, in comparison with the spectrum of coronene.
The lowest energy excitation of HBC (D1), corresponding to the Clar's $\alpha$ band, is found in our study at 2.84 eV (\ref{table1}).
This value is in excellent agreement with recent experimental results of HBC in gas phase \cite{kokk+08apjl} and in rare gas matrix \cite{roui+09jcp}: these studies assign the lowest energy excitation of HBC, at 2.86 eV and at 2.85 eV, respectively.
In Ref. \cite{roui+09jcp} also the position of the $p$ band (D2, 3.22 eV, see \ref{table1}) is reported at 3.30 -- 3.35 eV, depending on the embedding gas matrix.
Also for this excitation our results compare very well with the experiment within an error of 3$\%$.
The first active peak of HBC (P1) stems from our calculations at 3.82 eV (\ref{table1}), also in very good agreement with the experimental data at about 3.7 eV (error of 5$\%$) \cite{roui+09jcp}.

As a consequence of the preserved symmetry, the main optical features in all the examined hexagonal PAHs remain unchanged throughout the series (see \ref{fig2} and \ref{table1}). 
This is further confirmed by the resulting transition densities shown in \ref{fig3}.
In fact, the MO's contributing to the excitations are the same in all the structures, as their character is maintained by the molecular symmetry.
In the low energy spectral region, between D2 and P1, only one doubly degenerate excitation appears in the spectra of the larger molecules of this set (HBC, \ce{C114H30} and \ce{C222H42}).
As discussed for D1 and D2, also this dark excitation can be activated when coupled to vibrational modes (see e.g. Ref. \cite{roui+09jcp} for the case of HBC).
Higher energy excitations than P1 fall beyond the spectral range of interest (visible and near-UV bands) for coronene and HBC, while this is not the case for \ce{C114H30} and \ce{C222H42}.
As shown in \ref{fig2}, higher energy peaks stem from transitions between deeper occupied and higher unoccupied MOs, carrying much weaker intensity than P1.

%

\begin{figure}
\centering
\includegraphics[width=.95\textwidth]{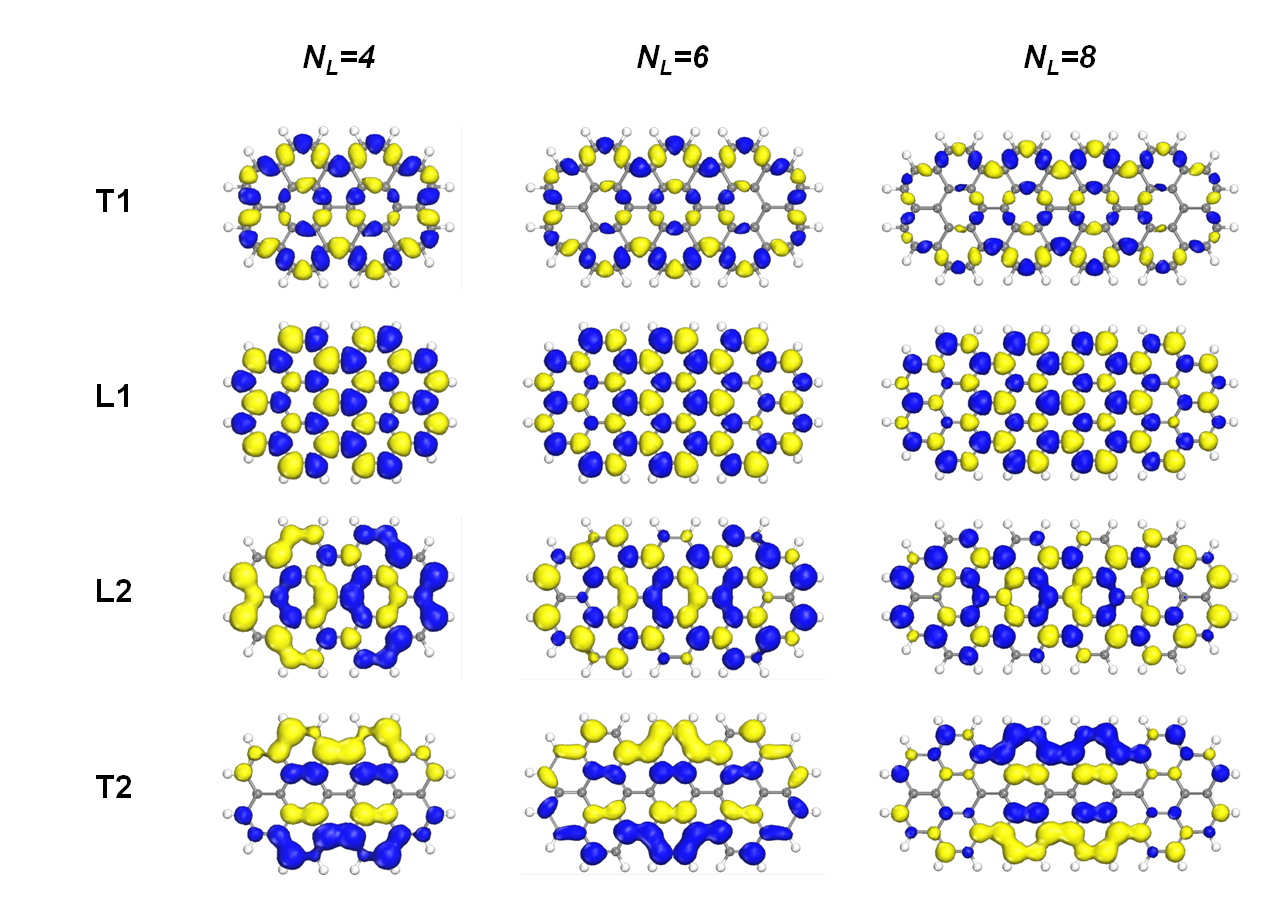}%
\caption{Transition density isosurfaces, computed as in \ref{eq:td}, of the examined optical excitations of ribbon-like GNFs with $D_{2h}$ point group symmetry.}
\label{fig5}
\end{figure}

Proceeding now with the results for the elongated GNFs, the drastic effect on the electronic and optical properties induced by symmetry lowering from $D_{6h}$ to $D_{2h}$ point group is evident already from the comparison between coronene and circumbiphenyl (see \ref{fig4}).
The doubly degenerate frontier orbitals of coronene (both HOMO and LUMO) are split in circumbiphenyl (HOMO and HOMO-1, LUMO and LUMO+1, \ref{fig4}a): the existence of two inequivalent axes ($x$ and $y$) in the molecular plane of the $N_L$=4 GNF induces a different parity for HOMO and HOMO-1, as well as for LUMO and LUMO+1.
In view of this reduced symmetry, the two lowest energy excitations (T1 and L1) are no longer dipole forbidden, as in the case of the $\alpha$ and $p$ bands in the hexagonal PAHs (\ref{fig4}b).
Moreover, the main peak P1 of the $D_{6h}$ series (Clar's $\beta$ band) splits into two distinct absorption lines, L2 and T2, with 0.1 eV energy separation.
These excitations are here labelled according to their polarization, as visualized in the transition density plots in \ref{fig5}: L1 and L2  are polarized along the longitudinal ($x$) axis of the flake,  while T1 and T2 along the transverse direction ($y$ axis).
By looking at the composition of the excitations (\ref{table2}), one notices that  T1 and T2 stem from the  HOMO $\rightarrow$ LUMO+1 and HOMO-1 $\rightarrow$ LUMO transitions, whereas L1 and L2 are due to the  HOMO $\rightarrow$ LUMO and HOMO-1 $\rightarrow$ LUMO+1 transitions.
However, while in T1 and T2 the weight of the two transitions is almost equivalent, the dominant contribution for L1 (L2) is given by HOMO $\rightarrow$ LUMO (HOMO-1 $\rightarrow$ LUMO+1). 
The anisotropy of circumbiphenyl is reflected also in the different oscillator strength of the longitudinally and transversally polarized excitations. 

The relative intensity of the excitations is connected with the anisotropy -- \textit{i.e.} length/width ratio -- of the $D_{2h}$ PAHs.
This phenomenon, which was discussed for longer flakes of the same family \cite{cocc+12jpcl}, can be definitely clarified here for these short structures.
As quantified in \ref{table2} and illustrated in \ref{fig4}b, going from the $N_L$=4 to the $N_L$=8 ribbon-like PAHs, the oscillator strengths increases by almost one order of magnitude, while the intensity of L2 is incremented only by a factor of 1.5 (see also \ref{table2}).
A visual picture of this feature is provided by the transition densities in \ref{fig5}, which indicates a similar modulation of L1 and L2 in the $N_L$=8 PAH, compared to the stronger dipolar character of L2 for circumbiphenyl.
In the case of the L1 excitation, the weight of the HOMO $\rightarrow$ LUMO increases with the length, at the expense of the HOMO-1 $\rightarrow$ LUMO+1 transition: in circumbiphenyl the ratio between these two single-particle transitions is about 1/3, while it decreases to less than 1/8 in the $N_L$=8 PAH.
The same occurs for L2, with a reversed contributions (\ref{table2}).
In the limit of very long ribbon-like PAHs with $N_W$=7 and length parameter up to $N_L$=20, corresponding to a length of about 88 \AA{}, L1 and L2 present an equal oscillator strength \citep{cocc+12jpcl}.
Also here the increasing length of the flakes induces a red shift of the spectra (see \ref{fig4}b), as observed in hexagonal PAHs (\ref{fig2}).
On the other hand, since the flake width is held fixed, the excitations with transverse polarization, T1 and T2, are very similar in all the considered $D_{2h}$ structures: their energy is only slightly modified, and the oscillator strength is preserved along with the relative weight of the dominant MO transitions, so that on longer flakes the direct excitons will be lower than the forbidden ones \cite{prez+08prb,yang+07nl}.
Finally, when comparing the energy of the first active excitation for symmetric ($D_{6h}$)  and elongated PAHs ($D_{2h}$), one notices that a peak at $\sim$2.9 eV is found for both the hexagonal \ce{C114H30} molecule and for the \ce{C52H20} flake ($N_L$=6), which present a striking ratio of molecular masses with respect to each other.

\begin{table}
\begin{tabular}{c|c|c|c|c|c}
\hline \hline
Molecule & Excitation & MO Transitions (weight) & $c_{\alpha \beta}^I$ & Energy [eV] & OS \\ \hline \hline
\multirow{8}{*}{$N_L$=4 (\ce{C38H16}) } & \multirow{2}{*}{T1} & H $\rightarrow$ L+1 (0.46) & -0.676 & \multirow{2}{*}{2.78} & \multirow{2}{*}{$\sim 10^{-4}$} \\
 & & H-1 $\rightarrow$ L (0.44) & 0.666 & & \\ \cline{2-6}
 & \multirow{2}{*}{L1} & H $\rightarrow$ L (0.71) & 0.842 & \multirow{2}{*}{3.14} & \multirow{2}{*}{0.17} \\
  & & H-1 $\rightarrow$ L+1 (0.25) & 0.502 & & \\ \cline{2-6}
 & \multirow{2}{*}{L2} & H $\rightarrow$ L (0.26) & 0.507 & \multirow{2}{*}{3.83} & \multirow{2}{*}{3.05} \\
  & & H-1 $\rightarrow$ L+1 (0.69) & -0.834 & & \\ \cline{2-6}
 & \multirow{2}{*}{T2} & H $\rightarrow$ L+1 (0.42) & 0.649 & \multirow{2}{*}{3.93} & \multirow{2}{*}{1.59} \\
  & & H-1 $\rightarrow$ L (0.43) & 0.660 & & \\ \cline{2-6}
  \hline \hline
  \multirow{8}{*}{$N_L$=6 (\ce{C52H20})} & \multirow{2}{*}{T1} & H $\rightarrow$ L+1 (0.44) & 0.663 & \multirow{2}{*}{2.66} & \multirow{2}{*}{$\sim 10^{-4}$} \\
 & & H-1 $\rightarrow$ L (0.42) & -0.646 & & \\ \cline{2-6}
 & \multirow{2}{*}{L1} & H $\rightarrow$ L (0.78) & 0.884 & \multirow{2}{*}{2.92} & \multirow{2}{*}{0.54} \\
  & & H-1 $\rightarrow$ L+1 (0.15) & 0.387 & & \\ \cline{2-6}
 & \multirow{2}{*}{L2} & H $\rightarrow$ L (0.16) & 0.403 & \multirow{2}{*}{3.57} & \multirow{2}{*}{4.04} \\
  & & H-1 $\rightarrow$ L (0.76) & -0.873 & & \\ \cline{2-6}
 & \multirow{2}{*}{T2} & H $\rightarrow$ L+1 (0.43) & 0.654 & \multirow{2}{*}{3.75} & \multirow{2}{*}{1.63} \\
  & & H-1 $\rightarrow$ L (0.45) & 0.671 & & \\ \cline{2-6}
  \hline \hline
  \multirow{7}{*}{$N_L$=8 (\ce{C66H24})} & \multirow{2}{*}{T1} & H $\rightarrow$ L+1 (0.41) & 0.643 & \multirow{2}{*}{2.59} & \multirow{2}{*}{$\sim 10^{-4}$} \\
 & & H-1 $\rightarrow$ L (0.39) & 0.627 & & \\ \cline{2-6}
 & L1 & H $\rightarrow$ L (0.80) & 0.895 & 2.80 & 1.04 \\ \cline{2-6}
 & \multirow{2}{*}{L2} & H $\rightarrow$ L (0.10) & 0.323 & \multirow{2}{*}{3.42} & \multirow{2}{*}{4.82} \\
  & & H-1 $\rightarrow$ L (0.77) & 0.880 & & \\ \cline{2-6}
 & \multirow{2}{*}{T2} & H $\rightarrow$ L+1 (0.41) & -0.639 & \multirow{2}{*}{3.65} & \multirow{2}{*}{1.57} \\
  & & H-1 $\rightarrow$ L (0.43) & 0.658 & & \\ \cline{2-6}
  \hline \hline
\end{tabular} 
\caption{Summary of the main optical excitations of rectangular PAHs ($D_{2h}$ point group symmetry), including energy, oscillator strength (OS) and composition in terms of molecular orbital transitions with the corresponding CI coefficients and weight. Note that only contributions larger than 10$\%$ have been included.
}
\label{table2}
\end{table}
%

\section{Conclusions}
In conclusion, we have presented a detailed quantum-chemical analysis of the effects of size and anisotropy on the optical properties of two prototypical classes of PAHs, with $D_{6h}$ and $D_{2h}$ symmetry.
Having ruled out spurious effects with the choice of structures with purely armchair edges, the optical properties of these systems 
appear strongly dependent both on the size and on the anisotropy of the molecule.
Most importantly, our findings indicate that the energy of the lowest energy active peak depends more crucially on the molecular symmetry, than on the molecular weight, related in this case to the number of C atoms in the structure.
In GNFs with $D_{2h}$ symmetry optically active peaks appear in the visible band already in the spectrum of circumbiphenyl (\ce{C38H16}).
On the other hand, only hexagonal molecules with a few hundreds of C atoms can absorb a photon in the same spectral region.
The results of this work provide significant insights into the interpretation of the experimental spectra of PAHs and consequently into the physical mechanisms driving their optical absorption.
Moreover, they confirm the large tunability of this class of compounds upon size and symmetry variations, regarding the optical properties.
In the field of optoelectronics these outcomes represent a valuable indication in view of designing molecular nanodevices.

Finally, the detailed investigation carried out in the present study on the size and anisotropy effects on the optical spectra of representative classes of PAHs can serve as a guide for the astronomical observation of these carbon compounds.
Specifically, the symmetry analysis of the main excitations in the visible region with respect of the molecular weight is a valuable indication for determining the relative abundance of medium-size aromatic molecules, and hence clarifying the composition of the ISM.

\begin{acknowledgement}
The authors are grateful to Stefano Corni and Franco Gianturco for stimulating discussions. CINECA is acknowledged for computational support. Part of this research was supported by: the Italian Ministry of Research through the national projects PRIN-GRAF (Grant No. 20105ZZTSE), FIRB-FLASHit (Grant No. RBFR12SWOJ), and the program ``Progetto Premiale 2012" - project ABNANOTECH; the Italian Ministry of Foreign Affairs through the Grant No. US14GR12. M. J. C. acknowledges support from FAPESP and CNPq (Brazil).
\end{acknowledgement}
\begin{tocentry}%
\centering
\includegraphics[width=9cm]{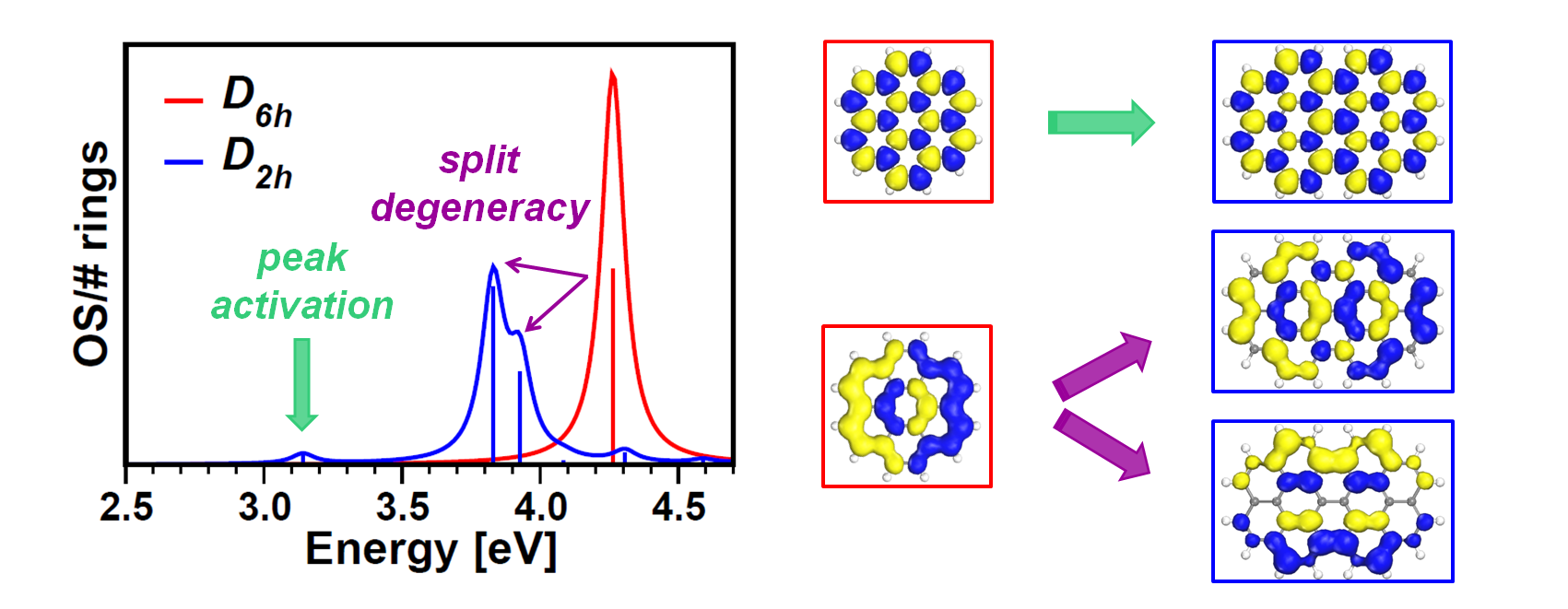}
\end{tocentry}
%
\providecommand*{\mcitethebibliography}{\thebibliography}
\csname @ifundefined\endcsname{endmcitethebibliography}
{\let\endmcitethebibliography\endthebibliography}{}



\begin{mcitethebibliography}{62}
\bibitem[Not()]{Note-2}
 We recently addressed this issue in the analysis of the optical spectrum of
  icosahedral \ce{C60}. A detailed discussion can be found in Ref.
  \cite{cocc+13jpcc}.\relax
\mciteBstWouldAddEndPunctfalse
\mciteSetBstMidEndSepPunct{\mcitedefaultmidpunct}
{}{\mcitedefaultseppunct}\relax
\EndOfBibitem
\bibitem[Not()]{Note-3}
 An exhaustive and systematic analysis of the size effects on the electronic
  and optical properties of a representative set of PAHs has been recently
  presented by Malloci and coworkers, see Ref. \cite{mall+11cp}.\relax
\mciteBstWouldAddEndPunctfalse
\mciteSetBstMidEndSepPunct{\mcitedefaultmidpunct}
{}{\mcitedefaultseppunct}\relax
\EndOfBibitem
\end{mcitethebibliography}
\end{document}